# One-Way Quantum Computer Simulation

Eesa Nikahd, Mahboobeh Houshmand, Morteza Saheb Zamani, Mehdi Sedighi

Quantum Design Automation Lab
Department of Computer Engineering and Information Technology
Amirkabir University of Technology
Tehran, Iran
{e.nikahd, houshmand, szamani, msedighi}@aut.ac.ir

Abstract— *In one-way quantum computation (1WQC) model, universal quantum computations are performed using measurements to designated qubits in a highly entangled state. The choices of bases for these measurements as well as the structure of the entanglements specify a quantum algorithm. As scalable and reliable quantum computers have not been implemented yet, quantum computation simulators are the only widely available tools to design and test quantum algorithms. However, simulating the quantum computations on a standard classical computer in most cases requires exponential memory and time. In this paper, a general direct simulator for 1WQC, called OWQS, is presented. Some techniques such as qubit elimination, pattern reordering and implicit simulation of actions are used to considerably reduce the time and memory needed for the simulations. Moreover, our simulator is adjusted to simulate the measurement patterns with a generalized flow without calculating the measurement probabilities which is called extended one-way quantum computation simulator (EOWQS). Experimental results validate the feasibility of the proposed simulators and that OWQS and EOWQS are faster as compared with the well-known quantum circuit simulators, i.e., QuIDDPro and libquantum for simulating 1WQC model*[1].

## 1 INTRODUCTION

Quantum information [2] is an interdisciplinary field which combines quantum physics, mathematics and computer science. Quantum computers have some advantages over the classical ones, e.g., they give dramatic speedups for tasks such as integer factorization [3] and database search [4]. They use the quantum mechanical phenomena such as superposition, measurement and entanglement [2]. Novel ideas have been introduced based on the use of measurement and entanglement to perform quantum computations, which are referred as measurement-based quantum computation (MBQC) [5]. The computational resources in measurement-based quantum computing models can be characterized by graph states. Graph states are special multi-qubit quantum states that can be shown by graphs in which each node represents a qubit and each edge represents an entanglement between pairs of qubits. Multi-qubit GHZ states with applications in quantum communication, or cluster states of arbitrary dimensions are some examples of graph states [6].

[1]This paper is an extended version of the paper presented at Euromicro DSD conference [1].

Two basic models of MBQC are teleportation quantum computation (TQC) and one-way quantum computation (1WQC), first proposed by Raussendorf and Briegel [7]. 1WQC has drawn researchers' attentions, mainly because it offers different and promising physical realizations of quantum computations.

In 1WQC, the quantum correlations in an entangled state, called cluster state or graph states, are utilized to perform quantum computations by single-qubit measurements only. The universality of two-dimensional cluster states has been proved [7]. The needed computations, specified in measurement patterns, are driven by irreversible projective measurements and hence, the model is called "one-way".

As scalable and reliable quantum computers have not been implemented yet, quantum computation simulators are the only widely available tools to design and test quantum algorithms. The most important challenge of the classical simulation of quantum computation models is the exponential time and memory complexity and the proposed simulators attempt to reduce such complexities.

Since the number of qubits in a 1WQC measurement pattern is considerably more than the equivalent quantum circuit, their simulation is more complex and the methods [8-12] proposed for simulating the quantum circuit model cannot be directly used for practical simulation of 1WQC, only by adding the capability of applying measurements in non-standard basis.

The aim of this paper is to overcome the circuit model simulators limitations in simulation of 1WQC patterns and present a general tool to directly simulate the 1WQC model. Two main techniques, namely qubit elimination and pattern reordering, are proposed to considerably reduce the state complexity as well as the time and memory needed for the simulations of 1WQC measurement patterns. After using these techniques, we can utilize previously proposed simulation techniques for the circuit model to achieve further optimizations for simulating the 1WQC model. Moreover, a proposed technique for implicit simulation of actions can be applied due to the limited number of basic actions in the 1WQC model. These three techniques are used to present an array-based 1WQC simulator called OWQS. In addition, OWQS is adjusted to simulate the measurement patterns with a generalized flow [13] without calculating the measurement probabilities which is called extended one-way quantum computation simulator (EOWQS). One of the applications of OWQS and EOWQS is the ability to answer to the recognition problem as stated in [14]. This problem is for recognizing what unitary a given 1WQC pattern implements.

The paper is organized as follows. In the next section, the preliminaries are presented. In Section 3, the related work is reviewed. In Section 4, the proposed approach is explained. Experimental results are presented in Section 5 and finally Section 6 concludes the paper with suggestions for future research.

## 2 Preliminaries

### *2.1 Qubits, quantum states and gates*

In classical binary computation, a bit assumes two distinct values, 0 and 1. Bits constitute the building blocks of the classical information theory. In an analogous manner, quantum bits or *qubits* are the fundamental units of information in quantum computing. A qubit is a unit vector in a two-dimensional Hilbert space, $\mathcal{H}_2$ whose basis vectors are denoted as:

$$|0\rangle = \begin{pmatrix} 1 \\ 0 \end{pmatrix} \text{and} |1\rangle = \begin{pmatrix} 0 \\ 1 \end{pmatrix}.$$

Unlike classical bits, qubits can be in a superposition of $|0\rangle$ and $|1\rangle$ represented by $|\psi\rangle = \alpha|0\rangle + \beta|1\rangle$ where $\alpha$ and $\beta$ are complex numbers such that $|\alpha|^2 + |\beta|^2 = 1$. If a measurement in the standard basis $\{|0\rangle, |1\rangle\}$, is applied to the state $|\psi\rangle$, the outcome will be 0(1) with the probability $|\alpha|^2$ ($|\beta|^2$), and the state immediately after the measurement is $|0\rangle$ ($|1\rangle$). The state of an $n$-qubit quantum system (quantum register) is represented by a column vector in a $2^n$-dimensional Hilbert space, $\mathcal{H}_2^n$ as follows:

$$|\Psi\rangle = |\psi_1 ... \psi_n\rangle = |\psi_1\rangle \otimes ... \otimes |\psi_n\rangle = \begin{pmatrix} \alpha_1 \\ \vdots \\ \alpha_{2^n} \end{pmatrix}$$

where $\otimes$ represents tensor product operation and $|\alpha_1|^2 + ... + |\alpha_{2^n}|^2 = 1$.

Entanglement is a unique quantum mechanical resource that plays a key role in many of the most interesting applications of quantum computation and quantum information. A multi-qubit quantum state $|\Psi\rangle$ is said to be entangled if it cannot be written as the tensor product $|\psi_1\rangle \otimes |\psi_2\rangle$ of two pure states. For example, the EPR pair $|\Psi\rangle = (|00\rangle + |11\rangle)/\sqrt{2}$ is an entangled quantum state.

There are a number of models for the evolution of quantum computation. The main model to explore quantum computation is the circuit model, based on unitary evolution of qubits by networks of gates. Every quantum gate is a linear transformation represented by a unitary matrix. A matrix $U$ is unitary if $UU^\dagger = I$, where $U^\dagger$ is the conjugate transpose of the matrix $U$. Since any unitary operation has an inverse, any quantum gate is reversible which means that given the state of a set of output qubits, it is possible to determine the state of its corresponding set of input qubits. Some useful single-qubit gates comprise Pauli set which are shown in the following:

$$I = \begin{bmatrix} 1 & 0 \\ 0 & 1 \end{bmatrix}, X = \begin{bmatrix} 0 & 1 \\ 1 & 0 \end{bmatrix}, Y = \begin{bmatrix} 0 & -i \\ i & 0 \end{bmatrix}, Z = \begin{bmatrix} 1 & 0 \\ 0 & -1 \end{bmatrix}.$$

Two other important single-qubit unitaries are Hadamard, $H$ and Phase gates, $P$:

$$H = \frac{1}{\sqrt{2}} \begin{bmatrix} 1 & 1 \\ 1 & -1 \end{bmatrix}, P = \begin{bmatrix} 1 & 0 \\ 0 & i \end{bmatrix}.$$

If $U$ is a gate that operates on a single qubit, then the *controlled-U* gate is a gate that operates on two qubits, i.e., control and target qubits, and $U$ is applied to the target qubit if the control qubit is $|1\rangle$ and otherwise, leaves it unchanged. For example, controlled-$Z$ (CZ) and controlled-NOT (CNOT) gates perform the $Z$ and $X$ operators respectively on the target qubit if the control qubit is $|1\rangle$ and no action is taken otherwise. The matrix representations of the CZ and CNOT gates are [2]:

$$\mathrm{CZ} = \begin{bmatrix} 1 & 0 & 0 & 0 \\ 0 & 1 & 0 & 0 \\ 0 & 0 & 1 & 0 \\ 0 & 0 & 0 & -1 \end{bmatrix}, \mathrm{CNOT} = \begin{bmatrix} 1 & 0 & 0 & 0 \\ 0 & 1 & 0 & 0 \\ 0 & 0 & 0 & 1 \\ 0 & 0 & 1 & 0 \end{bmatrix}$$

### *2.2 1WQC model*

The necessary computations in 1WQC are organized as patterns. A 1WQC pattern [14, 15] is defined as $P = (V, I, O, A)$, where $V$ is the set of qubits, $I \subset V$ the set of input qubits, $O \subset V$ the set of output qubits and $A$ a finite set of actions acting on $V$. The pattern is written as a sequence of four different types of actions. When performing a computation, the actions are applied from right to left. These actions are explained in the following.

- The *preparation action* $N_v$ prepares a qubit $v$ into state $|+\rangle = (|0\rangle + |1\rangle)/\sqrt{2}$ which is applied to all of the non-input qubits.
- *Entanglement action* $E_{uv}$ entangles qubits $u$ and $v$ by applying a CZ gate. The entanglement commands between the qubits can be represented by the edges in a graph, called the entanglement graph of the pattern.
- Single-qubit *measurement action* $M_v^\alpha$ measures the qubit $v$ in the orthonormal basis of:

$$|\pm\alpha\rangle = (|0\rangle \pm e^{-i\alpha}|1\rangle)/\sqrt{2} \tag{1}$$

  where $\alpha \in [0, 2\pi]$ is called the angle of measurement. The outcome of a measurement on a qubit $v$ is denoted by $s_v \epsilon \mathbb{Z}_2$. If the state of the qubit after the measurement is $|+\alpha\rangle$, then $s_v = 0$, and if it is $|-\alpha\rangle$, then $s_v = 1$. The measurement outcomes can be summed together modulo 2, which are called signals. A measurement can depend on other ones through two signals $s$ and $t$ as follows:

$${}^t\left[M_i^\alpha\right]^s = M_i^{(-1)^s \alpha + t\pi} \tag{2}$$

  To calculate the signals, all of the measurement results that appear in the signals $t$ and $s$ need to be known. This means that all those measurements must be performed before the dependent measurement.
- Single-qubit Pauli *correction actions* $X_v^s$ and $Z_v^s$ apply the Pauli $X$ and $Z$ gates to the qubit $v$, respectively, if $s$=1 and do nothing if $s$=0.

The set of actions $A$ must conform to the following rules:

*(D0) no action depends on an outcome not yet measured.*

*(D1) no action acts on a qubit already measured.*

*(D2) no action acts on a qubit not yet prepared unless it is an input qubit.*

*(D3) a qubit v is measured if and only if v is not an output.*

A pattern is in the standard form [14, 15] if all of the entanglement operations are at the beginning of the pattern, followed by all of the measurement operations and the correction operations are at the end of the pattern. The standard form is denoted by *CME* where *C, M* and *E* stand for correction, measurement and entanglement operators, respectively. The *quantum computation* depth of a pattern or just *quantum* depth is the depth of the execution of the pattern that is due to the dependencies of measurement and correction commands on previous measurement results. For example, the quantum depth of the standard pattern $P = \{\{1,2,\ldots,5\},\{1\},\{5\},X_5^{s_2+s_4}Z_5^{s_1+s_3}M_4^0\left[M_3^{\alpha}\right]^{s_2}\left[M_2^{\theta}\right]^{s_1}M_1^{\beta}E_{12345}\}$ is 4 due to the dependencies of the qubits 1235.

### *2.3 Projective measurement*

A set of mutually orthogonal projection operators $\{P_1, P_2, ..., P_m\}$ with the following properties constitute projective measurements [16]:

- $P_i^{\dagger} = P_i$
- $P_i^2 = P_i$
- $\sum_{i=1}^{m} P_i = I$

When this measurement is carried out in a system with state $|\psi\rangle$, then the result *i* is obtained with the probability:

$$prob_i = \langle\psi|P_i|\psi\rangle \tag{3}$$

and the state collapses to:

$$\frac{P_i|\psi\rangle}{\sqrt{prob_i}} \tag{4}$$

### *2.4 Generalized measurement*

The operators $\{M_1, M_2, ..., M_m\}$ on a Hilbert space $\mathcal{H}$ are called generalized measurement operators [16] if they satisfy:

$$\sum_{i=1}^{m} M_i^{\dagger} M_i = I \tag{5}$$

There are no conditions on $M_i$ other than this. When a generalized measurement with a set of measurement operators $\{M_1, M_2, ..., M_m\}$ is carried out in a system with state $|\psi\rangle$, then the result *i* is obtained with the probability:

$$prob_i = \langle\psi|M_i^{\dagger} M_i|\psi\rangle \tag{6}$$

and the state collapses to:

$$\frac{M_i|\psi\rangle}{\sqrt{prob_i}} \tag{7}$$

## 3 Related Work

In this section, a brief review of quantum circuit model simulators as well as 1WQC simulators is presented.

Viamontes et al. [9] define a new graph-based data structure for simulating quantum circuits called quantum information decision diagram (QuIDD). This data structure is used for the development of a quantum circuit simulator using BDD operations in the QuIDDPro software which uses the BDD software package CUDD. The main motivation of QuIDD is that vectors and matrices which arise in quantum computing, exhibit repeated structures. QuIDD utilizes these similarities in the vectors and matrices in order to reduce the memory and the run-time needed for the simulations.

In [11], a graph-based quantum circuit simulator based on quantum multiple-valued decision diagram (QMDD) structure is developed. The QMDD structure exploits the regular structure of the matrices which represent quantum circuits and gates.

In the fifth version of Feynman Program simulator [17], some features such as single-qubit measurements in the non-standard basis, projective and the generalized measurements are added which have been claimed to be useful for the MBQC simulation.

A class of quantum circuits with a restricted gate library can be efficiently simulated in polynomial time on classical computers. The results of Gottesman-Knill Theorem [2] and its recent improvement by Aaronson and Gottesman [18] apply only to circuits with stabilizer gates ─CNOT, Hadamard, Phase, Pauli gates and measurement of observables in the Pauli group─ and stabilizer input states. Aaronson and Gottesman's algorithm to simulate circuits with these gates is called CHP, whose time and space requirements scale only quadratically with the number of qubits [18]. In [19], CHP is improved in such a way that it requires time and space of $O(n \log n)$ where $n$ is the number of qubits by using graph states to represent the system state, which is called GraphSim.

$P$-blocked simulation is another technique which has been proposed in the quantum circuit simulation domain [30]. The basic idea of this approach is decomposing the states into smaller distinct sub-states, whenever possible. The typical way to pursue such separable states is employing entanglement metrics and contriving state representations whose size depend on these metrics. Such representation is called $p$-blocked if no subset of $p$+1 qubits are entangled. In other words, the set of all qubits is partitioned into $k$ blocks $B_1, B_2, \ldots, B_k$, where each block contains at most $p$ entangled qubits. Therefore, the state complexity grows with the number of entangled qubits instead of the total number of qubits.

Libquantum [10] is a C library for the simulation of quantum mechanics, with a special focus on quantum computing. It provides implementation of quantum registers, basic operations for register manipulation such as the Hadamard gate or the CNOT gate, measurement in the standard basis, etc.

In [20], a tool was presented to simulate the circuit and one-way quantum computation models in a parallel environment provided by PC workstations connected in a standard Ethernet network. The heart of the algorithm is the vector state transformation by a unitary matrix $U$. The most important challenge in this approach is how to divide the tasks involved in the simulation among different nodes while an efficient implementation of measurements and other actions have not been addressed.

In [21], a 1WQC emulator, based on formal measurement calculus [14, 15] was presented. The approach focuses on 1WQC and attempts to perform a faithful emulation of physical systems. It aims to carry out every step described by the theories of quantum mechanics and computations. The main drawback of this approach is that measurement and other actions have been implemented as straightforward matrix-vector multiplication.

In [22], it is shown that 1WQC patterns on non-universal one-dimensional cluster states can be simulated on classic computers efficiently, i.e., in polynomial time in the number of qubits. However, as these 1WQC patterns are not universal, a general simulator proposed for simulating one-way quantum computers cannot be of polynomial complexity.

A 1WQC simulation algorithm by contracting tensor networks was proposed by Markov and Shi [23] and it was shown that 1WQC can be efficiently simulated if its underlying graph has a small tree-width.

# 4 PROPOSED APPROACH

In this section, practical algorithms for simulation of 1WQC patterns on conventional computers using state vector representation are described. The proposed techniques can decrease the state space as well as time and memory complexities of simulation. The extent of reduction in the complexities depends on the structure of the underlying graph of entanglements and the measurement dependencies.

The structure of OWQS simulator is shown in Figure 1. First, it takes a pattern in the standard form as well as the state of the input qubits. Then, the pattern actions are reordered by an algorithm called, PROA[2]. Subsequently, the actions are applied in the new order by the simulator core. Finally, the state of the output qubits is reported.

## *4.1 State representation*

State vector representation is used to represent the system state. Initially, all of the qubit states are separate (except the input qubits which can be entangled), therefore the separate states (called sub-states) can be saved in distinct vectors while the system's total state is composed of these sub-states. Two different sub-states are combined only when the entanglement operation is applied to the corresponding qubits in these sub-states. Thus, in an $n$-qubit system, the memory complexity of storing the state space is $O\left(2^m\right)$, where $m$ is the number of qubits in the largest sub-state. The sub-state of an $m$-qubit system is saved in an array of size $2^m$ of complex numbers. For example, the representation of a three-qubit state $|\psi\rangle = \frac{1}{\sqrt{2}}\left[(1-i)|100\rangle - (1+i)|101\rangle + (1+i)|110\rangle + (-1+i)|111\rangle\right]$ is shown in Eq. 8.

$$|\psi\rangle = \frac{1}{2\sqrt{2}}\left[0,0,0,0,1-i,-1-i,1+i,-1+i\right]^{\mathrm{T}} \tag{8}$$

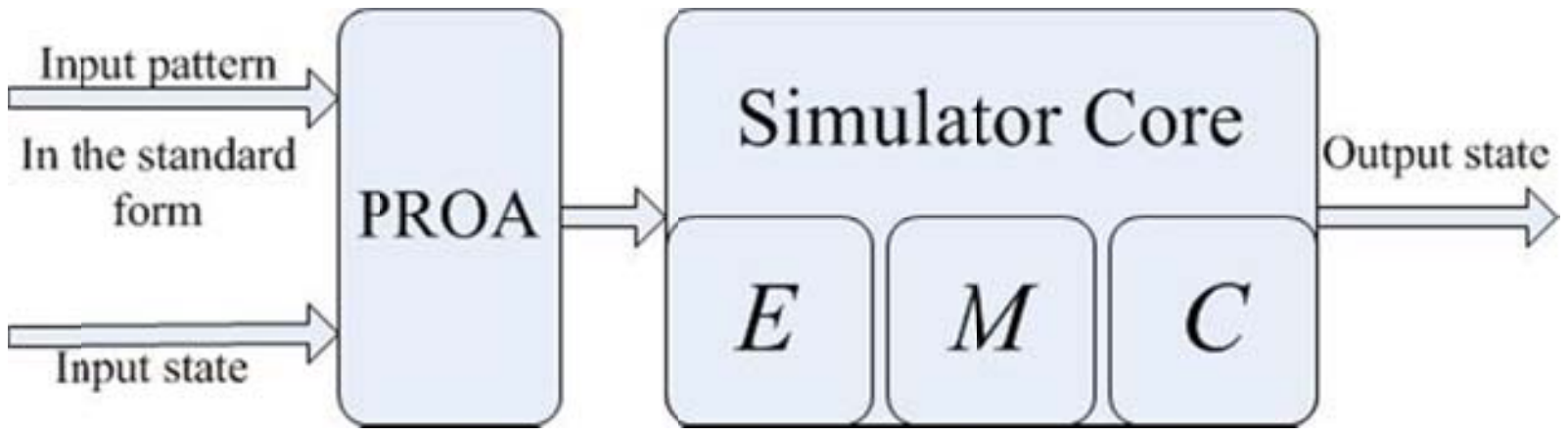

Figure 1.Structure of OWQS simulator where $E$, $M$ and $C$ stand for entanglement, measurement and correction, respectively.

[2]Pattern ReOrdering Algorithm

*4.2 Qubit elimination*

In 1WQC, single-qubit projective measurements are used and the measurement matrices $\{P_0, P_1\}$ are defined as:

$$P_0 = |+\alpha\rangle\langle+\alpha| = \frac{1}{2}\begin{bmatrix} 1 & e^{-i\alpha} \\ e^{i\alpha} & 1 \end{bmatrix} \tag{9}$$

$$P_1 = |-\alpha\rangle\langle-\alpha| = \frac{1}{2}\begin{bmatrix} 1 & -e^{-i\alpha} \\ -e^{i\alpha} & 1 \end{bmatrix} \tag{10}$$

However, in this paper a new generalized measurement basis is applied in order to implement *qubit elimination* technique easier as explained in the following.

Suppose that we are to perform a measurement on qubit $v$ in an $n$-qubit state $|\psi\rangle$. $|\psi\rangle$ can be divided into two parts based on $|0\rangle$ or $|1\rangle$ parts of the qubit $v$ which are defined by $\alpha_i$ and $\beta_i$, respectively as:

$$|\psi\rangle = \sum_{i=0}^{2^{n-1}-1} \alpha_i \, |b_{i,n} b_{i,n-1} \ldots 0_v \ldots b_{i,1}\rangle + \sum_{i=0}^{2^{n-1}-1} \beta_i \, |b_{i,n} b_{i,n-1} \ldots 1_v \ldots b_{i,1}\rangle$$

where $\alpha_i$ and $\beta_i$ are complex numbers and $(b_{i,n} b_{i,n-1} \ldots b_{i,1})$ is the binary expansion of number $i$. After each measurement, the state of the measured qubit $v$ is not important for further steps of simulation and can be removed from the state space. This is because in a 1WQC pattern each non-output qubit can be measured only once and after that no actions are allowed to apply on this qubit and finally only the state of output qubits determines final state of system. Therefore, the dimensionality of the $n$-qubit state $|\psi\rangle$ reduces from $2^n$ to $2^{n-1}$. In order to eliminate qubits after the measurement, we transform the known mentioned measurement basis $M(\alpha)$ to the standard basis as explained in the following.

Every single-qubit measurement can be associated with a unit vector on the Bloch sphere which corresponds to its +1 eigenstate and can be parameterized by the co-latitude θ and longitude φ of this vector, written as a pair of angles (θ, φ) (Figure 2) [24]. The measurement basis in 1WQC, that is $|\pm\alpha\rangle$ corresponds to the angles $(\pi/2, \alpha)$. In order to transform this measurement basis to the standard basis, it is necessary to make the measurement vector $|\pm\alpha\rangle$ coincide with the north pole of the Bloch sphere. Therefore, first, the measurement vector $|\pm\alpha\rangle$ is rotated around the $Z$-axis by an angle $-\alpha$ by applying $R_z(-\alpha) = \begin{bmatrix} 1 & 0 \\ 0 & e^{-i\alpha} \end{bmatrix}$ to the qubit. This action makes the vector coincide with $|+\rangle$. Then by applying a Hadamard gate to the qubit, the measurement vector will coincide with the north pole of the Bloch sphere and then we can measure this qubit in the standard basis. Figure 3 illustrates these steps.

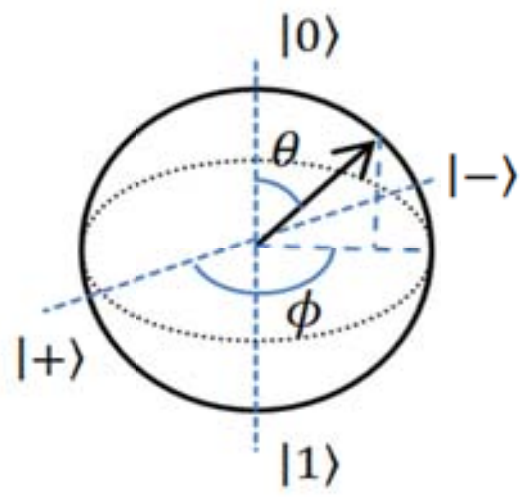

Figure 2. Single-qubit measurement can be represented by the pair of angles (θ, φ) on the Bloch sphere [24].

Therefore, the new generalized measurement basis is introduced as follows:

$$M_0 = |0\rangle\langle +\alpha| = \begin{bmatrix} 1 & 0 \\ 0 & 0 \end{bmatrix} .H. \begin{bmatrix} 1 & 0 \\ 0 & e^{-i\alpha} \end{bmatrix} = \frac{1}{\sqrt{2}} \begin{bmatrix} 1 & e^{-i\alpha} \\ 0 & 0 \end{bmatrix} = \left[ m_{ij} \right]_{i,j\in\{0,1\}} \tag{11}$$

$$M_1 = |1\rangle\langle -\alpha| = \begin{bmatrix} 0 & 0 \\ 0 & 1 \end{bmatrix} .H. \begin{bmatrix} 1 & 0 \\ 0 & e^{-i\alpha} \end{bmatrix} = \frac{1}{\sqrt{2}} \begin{bmatrix} 0 & 0 \\ 1 & -e^{-i\alpha} \end{bmatrix} = \left[ m'_{ij} \right]_{i,j\in\{0,1\}} \tag{12}$$

Applying the new generalized measurement on a qubit can be done in three steps:

1. The measurement operators $M_0$ and $M_1$ with respect to $\alpha$ are determined by using Eq.s 11 and 12.
2. The probability of the measurement result is calculated by using Eq. 6.
3. The state after the measurement is computed from Eq. 7.

It can be readily verified that $M_i^\dagger M_i = P_i \,;\ i \in \{0,1\}$ :

$$M_0^\dagger M_0 = \frac{1}{\sqrt{2}} \begin{bmatrix} 1 & 0 \\ e^{i\alpha} & 0 \end{bmatrix} . \frac{1}{\sqrt{2}} \begin{bmatrix} 1 & e^{-i\alpha} \\ 0 & 0 \end{bmatrix} = \frac{1}{2} \begin{bmatrix} 1 & e^{-i\alpha} \\ e^{i\alpha} & 1 \end{bmatrix} = P_0$$

$$M_1^\dagger M_1 = \frac{1}{\sqrt{2}} \begin{bmatrix} 0 & 1 \\ 0 & -e^{i\alpha} \end{bmatrix} . \frac{1}{\sqrt{2}} \begin{bmatrix} 0 & 0 \\ 1 & -e^{-i\alpha} \end{bmatrix} = \frac{1}{2} \begin{bmatrix} 1 & -e^{-i\alpha} \\ -e^{i\alpha} & 1 \end{bmatrix} = P_1$$

and therefore:

$$prob_i = \langle \psi | M_i^\dagger M_i | \psi \rangle = \langle \psi | P_i | \psi \rangle \tag{13}$$

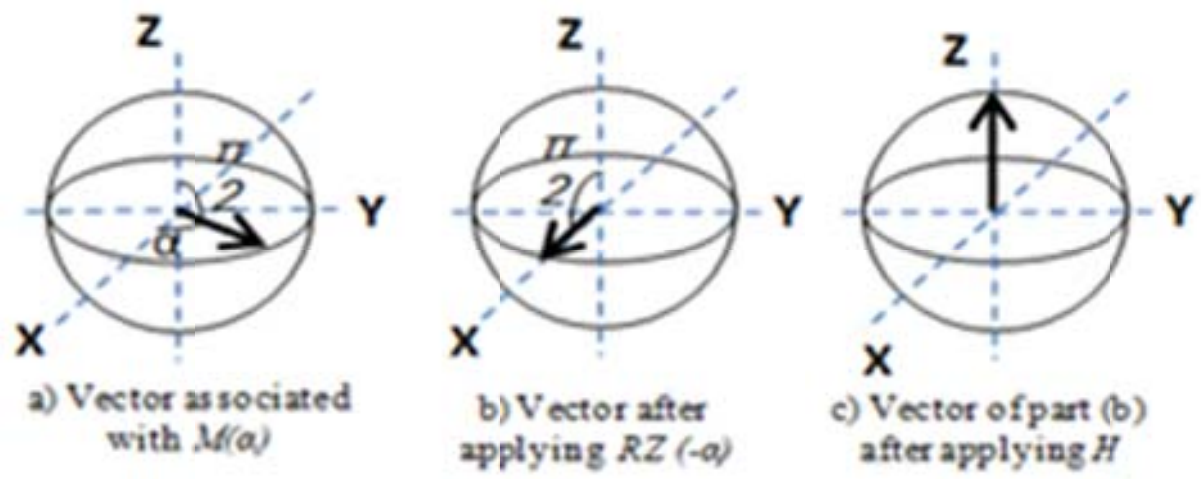

Figure 3. Steps to transform measurement basis $|\pm\alpha\rangle$ to the standard basis

By using this new basis, the measured qubit collapses to the state $|0\rangle$ or $|1\rangle$ instead of $|+\alpha\rangle$ or $|-\alpha\rangle$, respectively and the other qubits remain at the same states as for the measurement in the $|\pm\alpha\rangle$ basis. As a result, to eliminate measured qubit from the state space, simply half of the state with zero amplitudes is removed. Furthermore, while the traditional measurement operator has no zero element, the new proposed measurement matrix has only two non-zero elements, and this feature is utilized in implicit simulation of action in order to reduce the needed computations. Nevertheless, one can eliminate the measured qubit without using this new measurement operator but with more computational effort.

### *4.3 Pattern reordering*

Consider a pattern with size *n* in the standard form. All of the entanglement operations are performed at the beginning of the pattern. This leads to a state with size $2^n$. However, we propose an approach to keep the size of the states as small as possible by reordering the measurement actions and qubit elimination after each measurement to manage the state space as well as time and memory complexities.

*Proposition*: Before measuring a qubit *v* (with resolved dependencies) it is only needed to perform the entanglement operations to its adjacent qubits.

*Proof.* Without loss of generality, assume that the input pattern is in the standard form. The following relations hold:

a) The measurements on disjoint qubits can commute:

$$M_v^{\alpha_1} M_u^{\alpha_2} \Leftrightarrow M_u^{\alpha_2} M_v^{\alpha_1};\ u \neq v \tag{14}$$

b) The entanglements on qubits *(i,j)* and *(u,v)* commute:

$$E_{ij} E_{uv} \Leftrightarrow E_{uv} E_{ij} \tag{15}$$

c) The entanglement and the measurement operations on disjoint qubits commute:

$$M_{\vec{k}} E_{ij} \Leftrightarrow E_{ij} M_{\vec{k}};\ i, j \notin \vec{k} \tag{16}$$

where $\vec{k}$ is the set of qubits acted by *M* which does not contain *i* and *j*. Consider a subset of *CME* operations $C_{\vec{o}} M_{\vec{k}} M_v M_{\vec{t}} E_{\vec{s}} E_{v\vec{p}} E_{\vec{q}}$ and a target qubit *v* where the symbols as $\vec{x}$ represent the sets of qubits acted by *M* and *E* which do not include *v*. It should be mentioned that the measurements on the qubits on which *v* depends are assumed to be already performed. If we show that the measurement on *v* can be carried out only after performing the *E* actions on *v,* the proposition is proved. Using Eq. 15, all of the *v* entanglements can be moved to the right of the pattern. In this case, the pattern is as follows:

$$C_{\vec{o}} M_{\vec{k}} \boldsymbol{M}_v M_{\vec{t}} E_{\vec{s}} E_{\vec{q}} \boldsymbol{E}_{v\vec{p}}$$

Now by using Eq. 14, the pattern can be written as follows:

$$C_{\vec{o}} M_{\vec{k}} M_{\vec{t}} \boldsymbol{M}_v E_{\vec{s}} E_{\vec{q}} \boldsymbol{E}_{v\vec{p}}$$

Finally using Eq. 16, we will have:

$$C_{\vec{o}} M_{\vec{k}} M_{\vec{t}} E_{\vec{s}} E_{\vec{q}} \boldsymbol{M}_v \boldsymbol{E}_{v\vec{p}} \tag{17}$$

□

Therefore, using the above proposition, each qubit can be chosen in an appropriate order for measurement. Then, for the selected qubit, the entanglement operations with its adjacent qubits have to be performed, and finally the qubit can be measured. It is worth mentioning that in a pattern in the standard form, all of the correction commands are on the left of the pattern and remain unchanged. This is the basis of the pattern reordering algorithm. Since the qubits can be eliminated after each measurement, the complexity of the state space is kept small. Finding the best ordering of the measurements can be expressed as the following:

Figure 4 shows a graph *G* that represents the entanglement operations for the pattern of SWAP gate. Each vertex shows a qubit and the edges correspond to the entanglement operations. *q6* and *q8* are the output qubits which are not measured and the black vertices represent the qubits which have already been measured. The qubits which are ready to be measured are shown by white vertices.

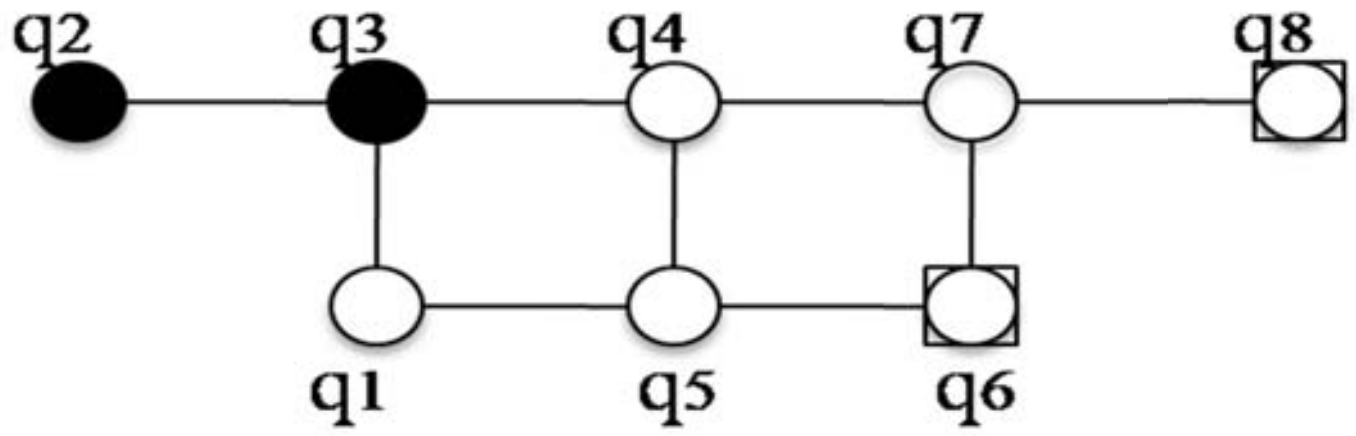

Figure. 4. SWAP entanglement graph. Each vertex represents a qubit. The black vertices and the white non-boxed vertices represent the measured qubits and the qubits ready to be measured, respectively. The output qubits which are not measured, are represented by the boxed vertices.

In Table I, an optimized order of qubits to be measured for the graph in Figure 4 is shown as an example. At stage 0, each qubit is in a separate set. Then *q2* is chosen for measurement, and therefore, the entanglement *E(q2,q3)* must be performed. To perform the entanglement, the sub-states containing the two qubits are tensored in a *two*-qubit state. After that, *q2* is measured and eliminated. The sub-states in the last column of the table represent the subsets after performing the measurements. At stage 2, *q3* is chosen for measurement. Then, *E(q3,q4)* and *E(q3,q1)* must be performed. Therefore, the sub-states containing *q3*, *q1* and *q4* are tensored. *q3* is measured and is then eliminated. At stage 3, *q1* is chosen and then *E(q1,q5)* is performed. *q1* is measured and then is deleted subsequently. The other qubits are chosen in a similar way, and finally, sub-state {*q6*, *q8*} shows the output state. This example indicates how the maximum size of state can be reduced from $2^8$ to $2^3$.

In the general case, dependent measurement cannot be performed as long as their dependencies are not resolved. Therefore, we define two measurement lists: *ready list* and *dependent list*. *Ready list* consists of the qubits with resolved dependencies and the qubits with dependent measurements are maintained in the *dependent list.* When all of the dependencies of a qubit have been removed, it is moved into the *ready list.*

To choose the best order of measurements, a heuristic approach is used. The first criterion is the size of the produced subset for measuring a qubit *v* (*MS*(*v*)) Higher priority is assigned to the smaller measurement state-spaces. The second criterion to be considered for each qubit *v* is the number of connections that it has to the output qubits (*OS*(*v*)). Less number of connections leads to a higher priority since the output qubits are never measured and will remain in the sub-state until the end of the simulation. Let $S(v)$ be the set of qubits which are dependent on the measurement result of qubit $v$. The size of $S(v)$ can be considered as the third criterion ($SS(v)$). Higher priority is assigned to the qubits whose size of $S$ is larger. Although selecting qubits in separate sub-states leads to some sub-states with sub-optimal size, when these sub-states need to be merged, a large sub-state will be produced. In order to prevent this, qubits which are connected to previously created sub-states are selected with a higher priority. This is the fourth criterion that is controlled by *flag* in Eq. 18 which assigns a cost to each unmeasured qubit.

$$Cost(v) = \begin{cases} \alpha * MS(v) + \beta * OS(v) - \gamma * SS(v) & flag = 1 \\ \alpha * ((MS(v)+1) * \delta) + \beta * OS(v) - \gamma * SS(v) & flag = 0 \end{cases} \tag{18}$$

where $\alpha$, $\beta$, $\gamma$ and $\delta$ are positive real numbers that control the importance of the first, second, third and fourth criteria, respectively and *flag* is set to 1 if qubit $v$ is connected to any previously selected qubits and set to 0 otherwise. In each step, the *Measurement state-space* of all qubits is generated, the qubit with the smallest cost is chosen, and then the *sub-states* are updated.

PROA consists of two nested loops. The number of iterations in the outer loop is equal to the number of qubits which are measured, i.e., $K = |V| - |O|$. In each iteration, one qubit is chosen to be measured. In the inner loop, for each candidate qubit $v$, the measurement state-space, which will be created in the case of choosing this qubit, is produced with a complexity proportional to the number of the pattern's qubits and the adjacent qubits to the qubit $v$. The number of the adjacent qubits to each qubit is $2E/|V|$ on average, where $E$ is the number of edges in the entanglement graph. Therefore, the time complexity to create measurement state-space for each candidate qubit is $\left(2E/|V|\right).|V| = 2E$ on average. In iteration $i$, the ready qubits to be measured is equal to $K$-$i$ in the worst case. Therefore, the total number of iterations is equal to:

$$\sum_{i=0}^{K} K - i = \sum_{i=1}^{K} i = \frac{K(K+1)}{2}$$

and the complexity of PROA algorithm is of

$$O\left(\frac{(K(K+1))}{2} * 2E\right) = O(K(K+1)E) = O(EK^2) \tag{19}$$

PROA aims to order actions in such a way that the problem state space is kept as small as possible. Whenever the quantum depth of the patterns is greater, PROA is more limited because of more measurement dependencies. Moreover, the entanglements of a selected qubit to its adjacent ones should be applied first and so a higher graph degree leads to larger sub-states. The output qubits are not measured and hence are not removed. Therefore, the number of adjacent qubits to each qubit and whether they are output qubits or not, affect PROA. The extent of the state complexity reduction cannot be exactly formulated but it is initiatively affected by the degree of the entanglement graph, the depth of the pattern and the number of output qubits. The smaller degree of entanglement graph, the less number of output qubits and the smaller depth of the pattern helps PROA reduce the state complexity more. Moreover, the values of $\alpha$, $\beta$, $\gamma$ and $\delta$ need to be appropriately set to utilize these characteristics, as explained in Section 5.

In an $n$-qubit system, the state space complexity of the problem is $O(2^m)$, where $m$ is the number of qubits in the largest sub-state. In the worst case, $m$ is equal to $n$ and in the best case for a pattern with a connected entanglement graph, $m$ is equal to $|O|+1$, where $|O|$ is the number of output qubits. The extent of reduction for each measurement pattern is shown in Table III, Section 5.

### 4.4 Implicit simulation of actions

In 1WQC, actions are limited to CZ, *X* and *Z* gates as well as single-qubit measurements. CZ, *X* and *Z* gates can be simulated with respect to their behavior, implicitly. Moreover, single-qubit measurement of a qubit in an *n*-qubit system leads to a measurement matrix with some regularity which is described in the following. This allows us to measure the specified qubit with no need to construct the measurement matrix and explicitly perform the matrix-vector multiplication.

Table I

An example of an optimized order of measurements

| Stage No. | Measurement orders | Measurement state-space | Sub-states |
|---|---|---|---|
| 0 | - | - | {1},{2},{3},{4},{5},{6},{7},{8} |
| 1 | 2 | {2,3} | {1},{3},{4},{5},{6},{7},{8} |
| 2 | 3 | {3,1,4} | {1,4},{5},{6},{7},{8} |
| 3 | 1 | {1,4,5} | {4,5},{6},{7},{8} |
| 4 | 4 | {4,5,7} | {5,7},{6},{8} |
| 5 | 5 | {5,6,7} | {6,7},{8} |
| 6 | 7 | {7,6,8} | {6,8} |

#### 4.4.1 Implicit simulation of entanglement, X and Z actions

In this section, the simulation of CZ, *X* and *Z* gates are explained. The basic idea is to simulate these gates with respect to their behavior instead of using the conventional matrix-vector multiplication, as explained in the following.

If the qubits *u* and *v* on which the CZ gate is to be applied, are in different sub-states, first the tensor product of the two corresponding sub-states are computed and then the CZ gate is applied. Let $|\psi\rangle$ be an *n*-qubit state that consists of the qubits *u* and *v* as follows:

$$|\psi\rangle=\sum_{i=0}^{2^{n-2}-1}\alpha_i\,|b_{i,n}b_{i,n-1}...0_u...0_v..b_{i,1}\rangle+\sum_{i=0}^{2^{n-2}-1}\beta_i\,|b_{i,n}b_{i,n-1}...0_u...1_v..b_{i,1}\rangle+\sum_{i=0}^{2^{n-2}-1}\gamma_i\,|b_{i,n}b_{i,n-1}...1_u...0_v..b_{i,1}\rangle+\sum_{i=0}^{2^{n-2}-1}\lambda_i\,|b_{i,n}b_{i,n-1}...1_u...1_v..b_{i,1}\rangle$$

where $\alpha_i,\beta_i,\gamma_i$ and $\lambda_i$ are complex numbers and $(b_{i,n}b_{i,n-1}..b_{i,1})$ is the binary expansion of *i*. In order to apply CZ, it is sufficient to change $\lambda_i$ into $-\lambda_i$ where $|u\rangle|v\rangle=|1\rangle|1\rangle$. Therefore, CZ can be done in $2^{n-2}$ steps. The pseudo code of this implementation is shown in Figure 5, which takes the*n*-qubit state *state* as well as *u* and *v* as the parameters. (In the code, the size of *state* is denoted by *size (state)*.)

As a result, one CZ gate is applied to the $u^{th}$ and $v^{th}$ qubits where $1\le v<u\le n$ and *λIndex* can be computed by only shift operations.

```
ENTANGLEMENT (state, u, v)
  for i ← 0 to size(state)/4
     do j ← i mod 2^(u−2)
     λIndex ← [j/2^(v-1)]*2^v + 2^(v-1) + (j mod 2^(v-1)) + [i/2^(u-2)] * 2^u + 2^(u-1)
     state[λIndex] ← state[λIndex] * (-1)
  end for
```

Figure 5. Pseudo code of entanglement action

Let $|\psi\rangle$ be an $n$-qubit state that consists of the qubit $v$ as follows:

$$|\psi\rangle = \sum_{i=0}^{2^{n-1}-1} \alpha_i \left|b_{i,n} b_{i,n-1} \ldots 0_v \ldots b_{i,1}\right\rangle + \sum_{i=0}^{2^{n-1}-1} \beta_i \left|b_{i,n} b_{i,n-1} \ldots 1_v \ldots b_{i,1}\right\rangle$$

In order to apply $X$ and $Z$ gates on the qubit $v$, it is sufficient to exchange $\alpha_i$ with $\beta_i$ for $X$ and change $\beta_i$ into $-\beta_i$ for $Z$ which can be applied both in $2^{n-1}$ steps.

*4.4.2 Implicit simulation of measurement*

Suppose that the measurement action $M(\alpha)$ is to be applied on the qubit $v$in an $n$-qubit state:

$$|\psi\rangle = \sum_{i=0}^{2^{n-1}-1} \alpha_i \left|b_{i,n} b_{i,n-1} \ldots 0_v \ldots b_{i,1}\right\rangle + \sum_{i=0}^{2^{n-1}-1} \beta_i \left|b_{i,n} b_{i,n-1} \ldots 1_v \ldots b_{i,1}\right\rangle$$

First, the measurement operator $M_i$ with respect to $\alpha$ is determined. In order to compute the state after the measurement, $M$ is calculated as follows:

$$M = \underbrace{I \otimes \ldots \otimes I}_{n\text{-}v} \otimes \underbrace{M_i \otimes I \otimes \ldots \otimes I}_{v} \tag{20}$$

Therefore, the straightforward computation of the state after the measurement using matrix-vector multiplication is of $O\left(2^{2n} * 2^n\right)$. In the following, the proposed approach to perform this is explained. The measurement operator $M$ has the following property:

- Each column and row of $M$ includes only the values of $M_i$ in the corresponding column and row as well as some other zero elements with certain regularities. In other words, there are only two non-zero elements in each column and row of $M$. For example:

$$M = I \otimes M_1 = \begin{bmatrix} 1 & 0 \\ 0 & 1 \end{bmatrix} \otimes \begin{bmatrix} m'_{00} & m'_{01} \\ m'_{10} & m'_{11} \end{bmatrix} = \begin{bmatrix} m'_{00} & m'_{01} & 0 & 0 \\ m'_{10} & m'_{11} & 0 & 0 \\ 0 & 0 & m'_{00} & m'_{01} \\ 0 & 0 & m'_{10} & m'_{11} \end{bmatrix}$$

$$M = I \otimes M_0 \otimes I = \begin{bmatrix} m_{00} & 0 & m_{01} & 0 & 0 & 0 & 0 & 0 \\ 0 & m_{00} & 0 & m_{01} & 0 & 0 & 0 & 0 \\ m_{10} & 0 & m_{11} & 0 & 0 & 0 & 0 & 0 \\ 0 & m_{10} & 0 & m_{11} & 0 & 0 & 0 & 0 \\ 0 & 0 & 0 & 0 & m_{00} & 0 & m_{01} & 0 \\ 0 & 0 & 0 & 0 & 0 & m_{00} & 0 & m_{01} \\ 0 & 0 & 0 & 0 & m_{10} & 0 & m_{11} & 0 \\ 0 & 0 & 0 & 0 & 0 & m_{10} & 0 & m_{11} \end{bmatrix}$$

Therefore, we only need to save the elements of $M_i$ (only four complex numbers) instead of maintaining the $2^{2n}$ complex numbers. Moreover, the *implicit* and *in-place matrix-vector multiplication* $M\left|\psi\right\rangle$ can be applied.

Implicit matrix-vector multiplication $M\left|\psi\right\rangle$ refers to applying $M\left|\psi\right\rangle$ without constructing matrix $M$ and passing up the zero elements during the multiplication. In-place matrix-vector multiplication refers to saving the results of the multiplication in the original vector without using any extra vectors. Let $M$ be a $2^n * 2^n$ matrix and $V$ a $2^n$-element column vector. Normally, applying $M*V$ needs to save the results in the new $2^n$-element column vector $V'$. However *in-place matrix-vector multiplication* can be used since there is only one non-zero element in each column of $M$. Therefore, each element of $V$ is used one time. Thus, we can use this element and then replace it with new one without saving the results in a new vector. These techniques are explained in the following example.

Let $\left|\psi\right\rangle = \alpha_0\left|000\right\rangle + \alpha_1\left|001\right\rangle + \alpha_2\left|100\right\rangle + \alpha_3\left|101\right\rangle + \beta_0\left|010\right\rangle + \beta_1\left|011\right\rangle + \beta_2\left|110\right\rangle + \beta_3\left|111\right\rangle$ be a three-qubit state and $M_v^\alpha$ be a measurement action on the second qubit ($v$=2). Assuming that the measurement outcome is zero, the measured state collapses to:

$$\left|\psi'\right\rangle = \frac{M_0\left|\psi\right\rangle}{\sqrt{prob_0}} = \frac{1}{\sqrt{2prob_0}} \begin{bmatrix} 1 & 0 & e^{-i\alpha} & 0 & 0 & 0 & 0 & 0 \\ 0 & 1 & 0 & e^{-i\alpha} & 0 & 0 & 0 & 0 \\ 0 & 0 & 0 & 0 & 0 & 0 & 0 & 0 \\ 0 & 0 & 0 & 0 & 0 & 0 & 0 & 0 \\ 0 & 0 & 0 & 0 & 1 & 0 & e^{-i\alpha} & 0 \\ 0 & 0 & 0 & 0 & 0 & 1 & 0 & e^{-i\alpha} \\ 0 & 0 & 0 & 0 & 0 & 0 & 0 & 0 \\ 0 & 0 & 0 & 0 & 0 & 0 & 0 & 0 \end{bmatrix} \begin{bmatrix} \alpha_0 \\ \alpha_1 \\ \beta_0 \\ \beta_1 \\ \alpha_2 \\ \alpha_3 \\ \beta_2 \\ \beta_3 \end{bmatrix}$$

$$= \frac{1}{\sqrt{2prob_0}} \begin{bmatrix} \alpha_0 + \beta_0 e^{-i\alpha} \\ \alpha_1 + \beta_1 e^{-i\alpha} \\ 0 \\ 0 \\ \alpha_2 + \beta_2 e^{-i\alpha} \\ \alpha_3 + \beta_3 e^{-i\alpha} \\ 0 \\ 0 \end{bmatrix} = \begin{bmatrix} \alpha'_0 \\ \alpha'_1 \\ 0 \\ 0 \\ \alpha'_2 \\ \alpha'_3 \\ 0 \\ 0 \end{bmatrix}$$

and after the elimination of the measured qubit, the final state will be:

$$|\psi''\rangle = \begin{bmatrix} \alpha'_0 \\ \alpha'_1 \\ \alpha'_2 \\ \alpha'_3 \end{bmatrix} \tag{21}$$

As a result, we need to move through half of the rows which only consist of $m_{00}$ and $m_{01}$ and then implicitly multiply these rows by $|\psi\rangle$ using only two multiplications. Each element of $|\psi\rangle$ is used only once for computing $\alpha'_i$, and so we can replace the old elements by the new ones (*in-place matrix-vector multiplication*). Computing the probability of the measurement is done implicitly too, as shown in Figure 6. Therefore, the measurement is performed by consuming constant size of extra memory and the time complexity of the algorithm for complex-number multiplications is $O\left(2^n\right)$. The pseudo code of this measurement operation is shown in Figure 6 which takes a measurement angle α, an *n*-qubit *state* and qubit number *v* as the input parameters. In this pseudo code, $p_{ij}$ are the elements of the former projective measurement matrix $P_0 = \begin{bmatrix} p_{00} & p_{01} \\ p_{10} & p_{11} \end{bmatrix}$ and $m_{ij}$ and $m'_{ij}$ are the elements of the new measurement matrices $M_0$ and $M_1$ respectively, and *rand*(*0,1*) returns a random number between zero and one.

### *4.5 Extended OWQS*

In this section, an accelerated extension of OWQS for a specific class of patterns is proposed. A pattern is *correct*, i.e., it implements a deterministic unitary if it is strongly deterministic. In patterns which have such this property, each branch occurs with the same probability *prob* = 0.5 and all of the branches implement an identical unitary (up to a global phase). Therefore, there is no need to compute the probabilities of the measurements. Browne et al. [13] present a necessary and sufficient condition for strong uniform determinism based on the geometry of the entanglement graph called generalized flow (*gflow*). Moreover, in [25] an algorithm is presented for finding a gflow in the patterns in polynomial time. If a pattern has a gflow, the probability of measurement results is 0.5; otherwise, the pattern is reported as an incorrect one. Furthermore, one may preselect the measurement results arbitrarily or simulate only the positive branch of the computations, in which all of the measurement outcomes are pre-selected to be zero. Therefore, the dependencies in the pattern are eliminated during the simulation and consequently, applying PROA algorithm potentially can lead to better results. This is because the results of all of the measurements are determined and therefore there is no dependent list and all of the qubits are ready for measurement from the beginning of simulation. This approach, which is called *EOWQS,* leads to a significant improvement in memory and run-time. The gflow algorithm [25] is incorporated into EOWQS and is applied before simulating each pattern as a preprocess operation.

**MEASUREMENT**(*state, α, v*)

*Create M and P matrices with respect to α*

*//computing the measurement probability of zero*

**for** $i \leftarrow 0$ to $size(state)$

**do** **if** $(i/2^{v-1} mod\ 2) = 0$

**then** $temp = p_{00} * state[i] + p_{01} * state[i + 2^{v-1}]$

**else**

**then** $temp = p_{10} * state[i + 2^{v-1}] + p_{11} * state[i]$

$prob_0 = prob_0 + conj\,(state[i] * temp)$

**end for**

*//computing the state after measurement*

**if** $(rand\,(0,1) \leq prob)$ *//if the measurement result is zero.*

**then** $a \leftarrow \dfrac{m_{00}}{\sqrt{prob_0}}$

$b \leftarrow \dfrac{m_{01}}{\sqrt{prob_0}}$

**else** $a \leftarrow \dfrac{m'_{10}}{\sqrt{1 - prob_0}}$ *//if the measurement result is one.*

$b \leftarrow \dfrac{m'_{11}}{\sqrt{1 - prob_0}}$

**for** $i \leftarrow 0$ to $size(state)/2$

**do** $j \leftarrow i\ mod\,(2^{v-1}) + [i/2^{v-1}] * 2^{v}$

$state[i] \leftarrow a * state[j] + b * state[j + 2^{v-1}]$

**end for**

Figure 6. Pseudo code for simulation of the measurement action

### *4.6 Efficient simulation*

For a cluster state of size *N*M* which is shown in Figure 7, our approach leads to $m=N+1$, where $N$ qubits on the right column of the cluster state are output qubits. To this end, the qubits for measurement are chosen column by column from left to right. In each column, qubits may be selected either from top to bottom or vice versa. Therefore, for a cluster state with $N \leq O(\log_2^n)$, the proposed approach is of polynomial complexity where $n$ is the number of qubits in the cluster state. In comparison to [22], the state space complexity of simulation of a cluster state 1**M* (1D cluster state) is O(1).

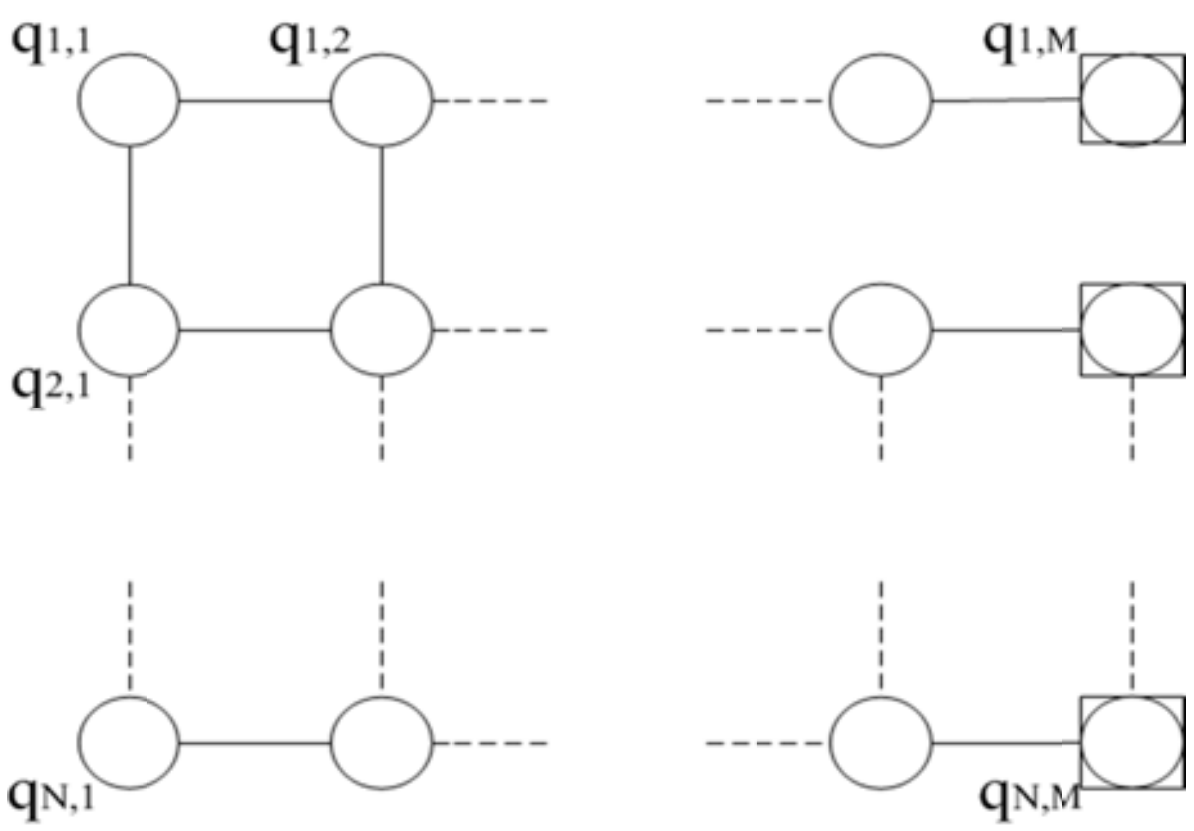

Figure 7. An $N*M$ cluster state. $N$ qubits in the boxes on the right most column are outputs.

### *4.7 A simple example*

In this section, the simulation of CNOT pattern with input state $|q2q1\rangle = |11\rangle$ is elaborated. The CNOT pattern is as follows [14]:

$$\text{CNOT} = \left(\{1,2,3,4\},\{1,2\},\{1,4\}, X_4^3 Z_4^2 Z_1^2 M_3^0 M_2^0 E_{13} E_{23} E_{34}\right) \tag{22}$$

OWQS takes this pattern as input and then the pattern is reordered by *PROA* algorithm as follows:

$$\text{CNOT} = \left(\{1,2,3,4\},\{1,2\},\{1,4\}, X_4^3 Z_4^2 Z_1^2 M_3^0 E_{13} E_{34} M_2^0 E_{23}\right) \tag{23}$$

The new reordered pattern is simulated by the *Simulator Core* of OWQS from right to left. The steps of applying the actions are illustrated in Figure 8. The initial state is shown in part (a). Note that total system state consists of four distinct sub-states. Then $E_{23}$ is applied. As a result, sub-states containing *q2* and *q3* are tensored (part (b)). *q2* is measured in the next step and then eliminated (part (c)). The other actions are performed in a similar way. Finally, the sub-state in part (f) shows the output state as $|q4q1\rangle = |01\rangle$.

## 5 Experimental Results

OWQS was implemented in C++ and was run on a workstation with 4GB RAM and Core2 Duo 2.6 GHz CPU. Some known quantum circuits were decomposed into CZ and $J(\alpha)$ gates. Then we applied the approach presented in [26, 27] in order to extract the corresponding pattern after performing the proposed optimizations which include standardization, signal shifting and Pauli simplifications.

In Table II, the run-time results obtained from our simulation algorithms are shown and compared with those produced by Emulator [21], QuIDDPro [9], and libquantum [10]. The numbers in the parentheses in the first column indicate the number of qubits used in the quantum circuit. Each pattern is simulated by all of the simulators with the same input states and the outputs are verified. We used a large sample of valid random input states in order to have a fair comparison. The table shows that the space complexity of the proposed approaches is considerably reduced by utilizing the *qubit elimination* and *pattern* *reordering* techniques. The speed-ups of OWQS and EOWQS compared with the QuIDDPro are shown in Figure 9 for the subset of benchmarks that the QuIDDPro can successfully simulates.

In [21], the space complexity of simulating an *n*-qubit pattern is $O(2^{2n})$, due to the $2^{2n}$-element projector $P$, and the patterns with at most 13 qubits can be simulated in this way. As all of the qubits in the standard 1WQC patterns are entangled at the beginning of simulation, in our benchmarks the libquantum fails to simulate patterns with more than 30 qubits.

The complexity of the proposed approach is $O(2^m)$, where $m$ is the number of qubits in the largest sub-state during the simulation.

| | |
|---|---|
| sub-state 1 | $\lvert q_1\rangle = (0 \quad 1)^{\mathrm{T}}$ |
| sub-state 2 | $\lvert q_2\rangle = (0 \quad 1)^{\mathrm{T}}$ |
| sub-state 3 | $\lvert q_3\rangle = (1 \quad 1)^{\mathrm{T}}/\sqrt{2}$ |
| sub-state 4 | $\lvert q_4\rangle = (1 \quad 1)^{\mathrm{T}}/\sqrt{2}$ |

a) Initial state with state $\lvert 11\rangle$ as input.

| | |
|---|---|
| sub-state 1 | $\lvert q_1\rangle = (0 \quad 1)^{\mathrm{T}}$ |
| sub-state 2 | $\lvert q_3 q_2\rangle = (0 \quad 1 \quad 0 \quad -1)^{\mathrm{T}}/\sqrt{2}$ |
| sub-state 3 | $\lvert q_4\rangle = (1 \quad 1)^{\mathrm{T}}/\sqrt{2}$ |

b) System state after applying $E_{23}$

| | |
|---|---|
| sub-state 1 | $\lvert q_1\rangle = (0 \quad 1)^{\mathrm{T}}$ |
| sub-state 2 | $\lvert q_3\rangle = (1 \quad -1)^{\mathrm{T}}/\sqrt{2}$ |
| sub-state 3 | $\lvert q_4\rangle = (1 \quad 1)^{\mathrm{T}}/\sqrt{2}$ |

c) System state after applying $M_2^0$ with a zero measurement result.

| | |
|---|---|
| sub-state 1 | $\lvert q_4 q_3 q_1\rangle = (0 \quad 1 \quad 0 \quad 1 \quad 0 \quad 1 \quad 0 \quad -1)^{\mathrm{T}}/2$ |

d) Applying $E_{13}$ $E_{34}$

| | |
|---|---|
| sub-state 1 | $\lvert q_4 q_1\rangle = (0 \quad 0 \quad 0 \quad 1)^{\mathrm{T}}$ |

e) System state after applying $M_3^0$ with a one measurement result.

| | |
|---|---|
| sub-state 1 | $\lvert q_4 q_1\rangle = (0 \quad 1 \quad 0 \quad 0)^{\mathrm{T}}$ |

f) Final state after applying $X_4^3 Z_4^2 Z_1^2$ .

Figure 8. Steps of simulation of CNOT gate

Table II
Simulation run-time of 1WQC patterns by OWQS and EOWQS compared with Emulator [21], libquantum [10] and QuIDDPro [9]

| Gate | #qubits in pattern (*n*) | *Emulator*(sec) | *Libquantum*(sec) | *QuIDDPro*(sec) | *OWQS*(sec) | *EOWQS*(sec) |
|---|---|---|---|---|---|---|
| CNOT (2) | 4 | 0.001 | 0.0004 | 0.014 | 0.0005 | 0.0004 |
| Toffoli (3) | 17 | N/A[3] | 0.051 | 0.196 | 0.003 | 0.003 |

[3]Not-applicable: the simulator exits with an error mostly caused by memory limitation

| | | | | | | |
|---|---|---|---|---|---|---|
| Fredkin (3) | 45 | N/A | N/A | 2.333 | 0.009 | 0.008 |
| GHZGate (20) | 39 | N/A | N/A | 9.543 | 0.880 | 0.330 |
| GHZGate (23) | 45 | N/A | N/A | 17.921 | 7.078 | 2.681 |
| GHZGate (25) | 49 | N/A | N/A | 43.250 | 30.818 | 10.864 |
| QFT (3) | 30 | N/A | 38.337 | 0.948 | 0.007 | 0.005 |
| QFT (5) | 90 | N/A | N/A | 64.700 | 0.054 | 0.050 |
| QFT (8) | 240 | N/A | N/A | >1 hour | 0.670 | 0.650 |
| QFT (9) | 306 | N/A | N/A | N/A | 1.353 | 1.298 |
| QFT (10) | 380 | N/A | N/A | N/A | 2.530 | 2.442 |
| 4_49 (4) [28] | 88 | N/A | N/A | 37.854 | 0.042 | 0.042 |
| nth_prime4_inc (4) [28] | 100 | N/A | N/A | 47.580 | 0.060 | 0.058 |
| hwb4 (4) [28] | 66 | N/A | N/A | 21.677 | 0.025 | 0.024 |
| ham7 (7) [28] | 129 | N/A | N/A | 222.63 | 0.147 | 0.138 |
| Shor code (9) | 103 | N/A | N/A | 299.751 | 0.103 | 0.098 |
| Toffoli_Staircase (13) [29] | 97 | N/A | N/A | 9.966 | 0.105 | 0.068 |
| Toffoli_Staircase (17) [29] | 129 | N/A | N/A | 19.801 | 1.152 | 0.326 |
| Toffoli_Staircase (21) [29] | 161 | N/A | N/A | 39.711 | 17.316 | 3.463 |
| rc_adder4 (16) [30] | 150 | N/A | N/A | 35.606 | 1.988 | 0.487 |
| rc_adder5 (20) | 188 | N/A | N/A | 55.504 | 30.185 | 5.061 |
| rd73 (10) [28] | 218 | N/A | N/A | 281.666 | 0.488 | 0.468 |
| rd84 (15) [28] | 323 | N/A | N/A | N/A | 3.960 | 2.870 |
| 6sym (10) [28] | 206 | N/A | N/A | N/A | 2.708 | 0.982 |
| gf2_4 (12) [28] | 242 | N/A | N/A | N/A | N/A | 3.166 |

As stated in Section 4.3 the result of the PROA algorithm is highly affected by the values of $\alpha$, $\beta$ and $\gamma$ in Eq. 18. The results reported in this section are obtained as described in the following empirical approach.

Initially $\alpha$, $\beta$, $\gamma$ and $\delta$ are set to $|O|, |O|, 0.5$ and $|O|$ respectively. Then in $|O|$ iterations, $\alpha$ is decreased by one and finally the value that produces the best result in all of the iterations is used. It should be noted that PROA is a polynomial-time algorithm and its run-time is negligible as compared with the main simulations run-time. Furthermore RPOA is applied for each pattern only once and can be considered as a preprocess operation. As shown in Table III, this approach leads to the minimum possible state-space in many cases, i.e., $m = |O| + 1$ (the numbers indicated by*).

Table III
Simulation state-space of OWQS and EOWQS (O($2^m$)). * shows PROA leads to the best possible results of *m,* i.e., $|O|+1$

| Gate | *N* | *m*(OWQS) | *m* (EOWQS) |
|---|---|---|---|
| CNOT (2) | 4 | 3* | 3* |
| Toffoli (3) | 17 | 4* | 4* |
| Fredkin (3) | 45 | 4* | 4* |
| GHZGate (25) | 49 | 26* | 26* |
| QFT (3) | 30 | 4* | 4* |
| QFT (5) | 90 | 6* | 6* |
| QFT (8) | 240 | 9* | 9* |
| QFT (9) | 306 | 10* | 10* |
| QFT (10) | 380 | 11* | 11* |
| 4_49 (4) [28] | 88 | 6 | 6 |
| nth_prime4_inc (4) [28] | 100 | 5* | 5* |
| hwb4 (4) [28] | 66 | 5* | 5* |
| ham7 (7) [28] | 129 | 9 | 9 |
| Shor code (9) | 103 | 14 | 12 |
| Toffoli_Staircase (13) [29] | 97 | 15 | 15 |
| Toffoli_Staircase (17) [29] | 129 | 19 | 19 |
| Toffoli_Staircase (21) [29] | 161 | 23 | 23 |
| rc_adder4 (16) [30] | 150 | 20 | 17* |
| rc_adder5 (20) | 188 | 25 | 21* |
| rd73 (10) [28] | 218 | 12 | 12 |
| rd84 (15) [28] | 323 | 20 | 20 |
| 6sym (10) [28] | 206 | 22 | 16 |
| gf2_4 (12) [28] | 242 | 27 | 19 |

Due to the elimination of the qubit dependencies, EOWQS usually leads to a smaller state space than OWQS. Moreover, due to the lack of need for calculating probabilities, the simulation run-time and space complexity of EOWQS is reduced.

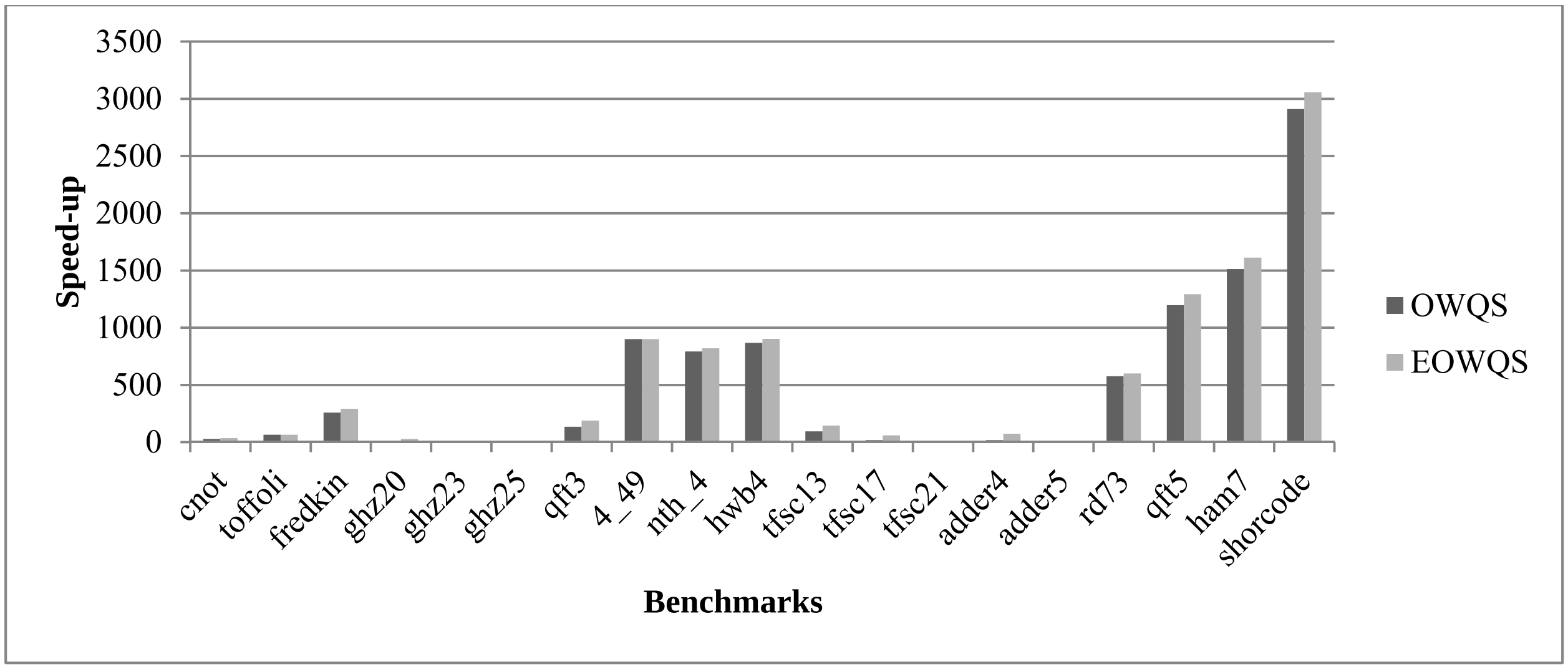

Figure 9. Speed-ups of OWQS and EOWQS compared with QuIDDPro

### *5.1 Results analyses and comparisions with previous quantum circuit model simulators*

As Table II and Figure 9 indicate, the methods proposed for simulating the quantum circuit model cannot be directly used for simulation of 1WQC in practice, as the number of qubits in a 1WQC pattern is considerably more than the equivalent quantum circuit.

This paper overcomes this problem, mainly by proposing *pattern reordering* and *qubit elimination* techniques. The proposed simulators do not use the previous quantum circuit model ideas. However, after using *pattern reordering* and *qubit elimination* techniques, other previously proposed circuit model simulation techniques can be exploited for obtaining further improvements.

The *p*-blocked simulation technique decomposes states into smaller states during the simulation. For example, assume that two qubits are entangled and hence they compose a single two-qubit entangled state. Applying some gates to them may change the state of the qubits into non-entangled ones. Therefore, they can be decomposed into two single-qubit non-entangled states. However, our simulators simply maintain qubits in the separated states at the beginning of simulation and do not decompose them during the simulation. The *p*-blocked technique can be used along with our proposed simulators for obtaining further improvements by decomposing each sub-state into the smaller ones whenever possible during the simulation. The extent of the improvement induced by the *p*-blocked simulation highly depends on the state of the input qubits and the applied gates (or the graph structure of the 1WQC pattern) where these determine the number of entangled qubits during the simulation.

QuIDDPro and QMDD are powerful graph-based circuit model simulators. The idea of using graph structure in this scope was born out of the similarity and repeated structure of matrices and vectors in quantum computing. Depending on the extent of regularity and similarity in the state vectors and matrices (gates), the state complexity varies from $O(1)$ to $O(2^n)$ while the worst case does not usually happen. However, our simulator simply uses an array structure to store states and matrices. Adding the graph structure to our simulators can also improve the results. Since the graph-based simulators utilize similarity and repeated structure of matrices and vectors, their effectiveness highly depends on the state of input qubits and applied gates (or graph structure of 1WQC pattern). The input states which lead to more similarity in the system state are preferred by graph-based simulators. The effectiveness of libquantum also depends on the state of input qubits because it benefits from the sparse structure for storing state vectors.

The GraphSim and the CHP simulators have both been proposed for a special sub-class of quantum circuits, i.e., stabilizer circuits whose simulation can be done in polynomial time according to Gottesman-Knill theorem. These simulators can be applied to 1WQC patterns only when the measurements are done in $X$, $Y$ and $Z$ bases and the input states are stabilizer states. Otherwise, these simulators are unable to simulate the patterns. However, our proposed simulators are general in the sense that they accept any kind of patterns with any measurement angles and arbitrary input states. As Table IV shows, the GraphSim simulator can be applied on a limited sub-set of 1WQC patterns of Table II and as expected, leads to smaller run-times for these patterns as compared with the proposed simulators. It is suggested to have GraphSim and the proposed simulators in a comprehensive simulation tool and simulate the stabilizer circuits using GraphSim.

Our simulators do not use any of the above factors or any other factors making the performance dependent on the input state and hence they do not depend on the state of input qubits. The simulation results also confirm this statement.

In order to verify the dependencies of the previously proposed quantum circuit and the proposed simulators on the input data, Table V compares the run-time of these approaches for different types of input data, i.e., when the input qubits are in the basis ($\{|0\rangle, |1\rangle\}$) or $|\pm\rangle$ states. As this table indicates, the run-time of OWQS and EOWQS do not depend on the state of input qubits, while the run-times of QuIDDPro and libquantum do.

Table IV

Simulation run-time of OWQS and EOWQS compared with the GraphSim simulator

| Gate | No. of qubits in pattern ($n$) | GraphSim(sec) | OWQS (sec) | EOWQS (sec) |
|---|---|---|---|---|
| CNOT | 4 | 0.0007 | 0.0005 | 0.0004 |
| SWAP | 8 | 0.0009 | 0.0014 | 0.0009 |
| GHZGate (20) | 39 | 0.0086 | 0.880 | 0.330 |
| GHZGate (23) | 45 | 0.0095 | 7.078 | 2.681 |
| GHZGate (25) | 49 | 0.0101 | 30.818 | 10.864 |

Table V

Simulation run-times of 1WQC patterns by OWQS and EOWQS compared with libquantum and QuIDDPro [9] for the standard basis ($\{|0\rangle, |1\rangle\}$) and $|\pm\rangle$ input state of each qubit.

| Gate | #qubits in pattern ($n$) | Input state | Emulator(sec) | Libquantum (sec) | QuIDDPro (sec) | OWQS (sec) | EOWQS(sec) |
|---|---|---|---|---|---|---|---|
| CNOT (2) | 4 | basis | 0.001 | 0.0003 | 0.013 | 0.0005 | 0.0004 |
| | | $|\pm\rangle$ | 0.001 | 0.0004 | 0.016 | 0.0005 | 0.0004 |
| Toffoli (3) | 17 | basis | N/A | 0.042 | 0.172 | 0.003 | 0.003 |
| | | $|\pm\rangle$ | N/A | 0.057 | 0.324 | 0.003 | 0.003 |

| | | | | | | | |
|---|---|---|---|---|---|---|---|
| Fredkin (3) | 45 | basis | N/A | N/A | 1.892 | 0.009 | 0.007 |
| | | $\|\pm\rangle$ | N/A | N/A | 7.296 | 0.010 | 0.008 |
| GHZGate (20) | 39 | basis | N/A | N/A | 0.260 | 0.871 | 0.330 |
| | | $\|\pm\rangle$ | N/A | N/A | N/A | 0.882 | 0.335 |
| GHZGate (23) | 45 | basis | N/A | N/A | 0.336 | 7.080 | 2.679 |
| | | $\|\pm\rangle$ | N/A | N/A | N/A | 7.081 | 2.688 |
| GHZGate (25) | 49 | basis | N/A | N/A | 0.405 | 30.837 | 10.850 |
| | | $\|\pm\rangle$ | N/A | N/A | N/A | 30.818 | 10.964 |
| QFT (3) | 30 | basis | N/A | 34.211 | 0.732 | 0.007 | 0.005 |
| | | $\|\pm\rangle$ | N/A | 41.642 | 2.061 | 0.008 | 0.005 |
| QFT (5) | 90 | basis | N/A | N/A | 36.129 | 0.054 | 0.052 |
| | | $\|\pm\rangle$ | N/A | N/A | >1 hour | 0.061 | 0.053 |
| QFT (8) | 240 | basis | N/A | N/A | >1 hour | 0.660 | 0.654 |
| | | $\|\pm\rangle$ | N/A | N/A | >1 hour | 0.663 | 0.652 |
| QFT (9) | 306 | basis | N/A | N/A | >1 hour | 1.340 | 1.301 |
| | | $\|\pm\rangle$ | N/A | N/A | N/A | 1.361 | 1.308 |
| QFT (10) | 380 | basis | N/A | N/A | N/A | 2.521 | 2.427 |
| | | $\|\pm\rangle$ | N/A | N/A | N/A | 2.539 | 2.444 |
| 4_49 (4) [28] | 88 | basis | N/A | N/A | 23.244 | 0.043 | 0.041 |
| | | $\|\pm\rangle$ | N/A | N/A | 323.910 | 0.042 | 0.042 |
| nth_prime4_inc (4) [28] | 100 | basis | N/A | N/A | 44.064 | 0.061 | 0.58 |
| | | $\|\pm\rangle$ | N/A | N/A | 420.050 | 0.061 | 0.060 |
| hwb4 (4) [28] | 66 | basis | N/A | N/A | 17.818 | 0.025 | 0.024 |
| | | $\|\pm\rangle$ | N/A | N/A | 69.991 | 0.025 | 0.023 |
| ham7 (7) [28] | 129 | basis | N/A | N/A | 173.333 | 0.147 | 0.140 |
| | | $\|\pm\rangle$ | N/A | N/A | N/A | 0.147 | 0.138 |
| Shor code (9) | 103 | basis | N/A | N/A | 289.664 | 0.100 | 0.095 |
| | | $\|\pm\rangle$ | N/A | N/A | 598.437 | 0.113 | 0.098 |
| Toffoli_Staircase (13) [29] | 97 | basis | N/A | N/A | 7.822 | 0.103 | 0.061 |
| | | $\|\pm\rangle$ | N/A | N/A | >1 hour | 0.110 | 0.073 |
| Toffoli_Staircase (17) [29] | 129 | basis | N/A | N/A | 15.315 | 1.143 | 0.333 |
| | | $\left\| \pm \right\rangle$ | N/A | N/A | >1 hour | 1.152 | 0.338 |
| Toffoli_Staircase (21) [29] | 161 | basis | N/A | N/A | 34.865 | 17.315 | 3.464 |
| | | $\left\| \pm \right\rangle$ | N/A | N/A | N/A | 17.327 | 3.473 |
| rc_adder4 (16) [30] | 150 | basis | N/A | N/A | 29.321 | 2.002 | 0.480 |
| | | $\left\| \pm \right\rangle$ | N/A | N/A | N/A | 2.008 | 0.493 |
| rc_adder5 (20) | 188 | basis | N/A | N/A | 47.677 | 30.178 | 5.060 |
| | | $\left\| \pm \right\rangle$ | N/A | N/A | N/A | 30.185 | 5.061 |
| rd73 (10) [28] | 218 | basis | N/A | N/A | 237.501 | 0.488 | 0.460 |
| | | $\left\| \pm \right\rangle$ | N/A | N/A | N/A | 0.489 | 0.471 |
| rd84 (15) [28] | 323 | basis | N/A | N/A | N/A | 3.960 | 2.871 |
| | | $\left\| \pm \right\rangle$ | N/A | N/A | N/A | 3.967 | 2.872 |
| 6sym (10) [28] | 206 | basis | N/A | N/A | >1 hour | 2.708 | 0.982 |
| | | $\left\| \pm \right\rangle$ | N/A | N/A | N/A | 2.711 | 0.982 |
| gf2_4 (12) [28] | 242 | basis | N/A | N/A | N/A | N/A | 3.162 |
| | | $\left\| \pm \right\rangle$ | N/A | N/A | N/A | N/A | 3.173 |

## 6 CONCLUSIONAND FUTURE WORKS

1WQC has drawn considerable attentions, mainly because it offers different physical realizations of the quantum computations. However, to the best of our knowledge no practical tool has been proposed for the simulation of this model. It should be mentioned that the number of qubits in the 1WQC pattern is considerably more than the equivalent quantum circuit which makes their simulation more complex. Therefore, conventional circuit model simulators such as QuIDDPro cannot be directly used for the simulation of 1WQC. In this paper, a practical approach to simulating 1WQC patterns on the classical computers, called OWQS, was proposed and then was extended in a way that it can reduce run-time and potentially space complexity of simulation by exploiting the concept of gflow. Two main techniques, *qubit elimination* and *pattern reordering* were presented to considerably reduce the state complexity as well as the time and memory needed for the simulations of 1WQC patterns. After using these techniques one can utilize previously proposed circuit model simulation ideas to achieve further optimizations. Using graph structure to represent system states [9] instead of arrays is underway.